\documentclass[runningheads,a4paper]{llncs}

\usepackage{latexsym,amssymb,amsmath}
\usepackage{epsf}
\usepackage{graphicx}
\usepackage{subfigure}

\begin{document}

\title{A Semantics-Aware Editing Environment\\ for Prolog in Eclipse}
\author{Jens Bendisposto \and Ian Endrijautzki \and Michael Leuschel \and David Schneider}
\institute{ 
 Softwaretechnik und Programmiersprachen\\
  Institut f\"ur Informatik\\
 Universit\"{a}t D\"{u}sseldorf\\  
 {\tt \{bendisposto,leuschel\}@cs.uni-duesseldorf.de} 
}
\tocauthor{Jens Bendisposto, Ian Endrijautzki, Michael Leuschel, and David Schneider
(University of D\"usseldorf)}

\date{}

\maketitle
\setcounter{page}{18}
\vspace*{-0.3cm}

\section{Introduction}


The open-source Eclipse platform has become hugely popular as an integrated development environment (IDE) for Java, mainly because of its editing features. The Java IDE comes with code highlighting, hover information, code completion, quick fixes and many more. It supports the development process by highlighting problematic aspects of the code and providing a convenient and robust way to improve it using refactorings.
All those features have a common goal: let the user stay focused on his task, writing code.

It would be very useful for the logic programming community to obtain an IDE with similar features for the development of Prolog programs.

However, writing an industrial strength integrated development environment like Eclipse for another language from scratch is a very difficult and complex task.
Fortunately, in previous work
 we have developed BE4 \cite{BeLeu_B07}, a framework for building semantic aware editors built on top of Eclipse and the Rigorous Open Development Environment for Complex Systems (RODIN).
BE4 contains a reliable, multi phase build process that can be used for arbitrary languages.
The build process was designed with the purpose of integrating different tools, even proprietary compilers into BE4. In particular, it is not restricted to Java. BE4 also contains a toolset to implement state-of-the-art editors, providing features such as syntactic and semantic highlighting, code completion, outline view, hover information, quick fixes and semantic checks.
Finally BE4 already comes with language plug-ins for classical B, CSP, Promela and SableCC, and a such its applicability
 has already been tested on a variety of source languages.

In this paper we present a Prolog plugin for Eclipse based upon BE4,
 and providing many features such as semantic-aware syntax highlighting,
  outline view, error marking, content assist, hover information, documentation generation, and quick fixes.
The plugin makes use of a Java parser for full Prolog with an integrated Prolog engine,
 and can be extended with further semantic analyses, e.g., based on abstract interpretation.


\section{Features}

Beside the obligatory syntax highlighting, our tool ProClipse offers a wide variety of features
 helping the user develop and adapt his or her Prolog programs.
Our features are based on a full syntactical and semantical analysis (i.e., they are not
 derived on approximate solutions using regular expressions).
 %
%
Thus far, the following features have been implemented:

\paragraph{Outline View:} ProClipse creates an outline view which represents an overview of
 the Prolog file or module, containing
  exported or non-exported predicates, defined DCG predicates and import directives.
Each item in the outline view can be used to quickly access the respective lines of the Prolog code.

\paragraph{Error presentation:} Syntax and semantic errors are highlighted 
in the editor view. A wiggly line underlines the erroneous part
of the source code and each line containing an error is tagged with an error marker.
The problems view lists all errors of all Prolog files in an Eclipse Prolog project
and can also be used to directly recall an erroneous source code line.
 
\begin{figure}[h]
\centering
\includegraphics[scale=0.31]{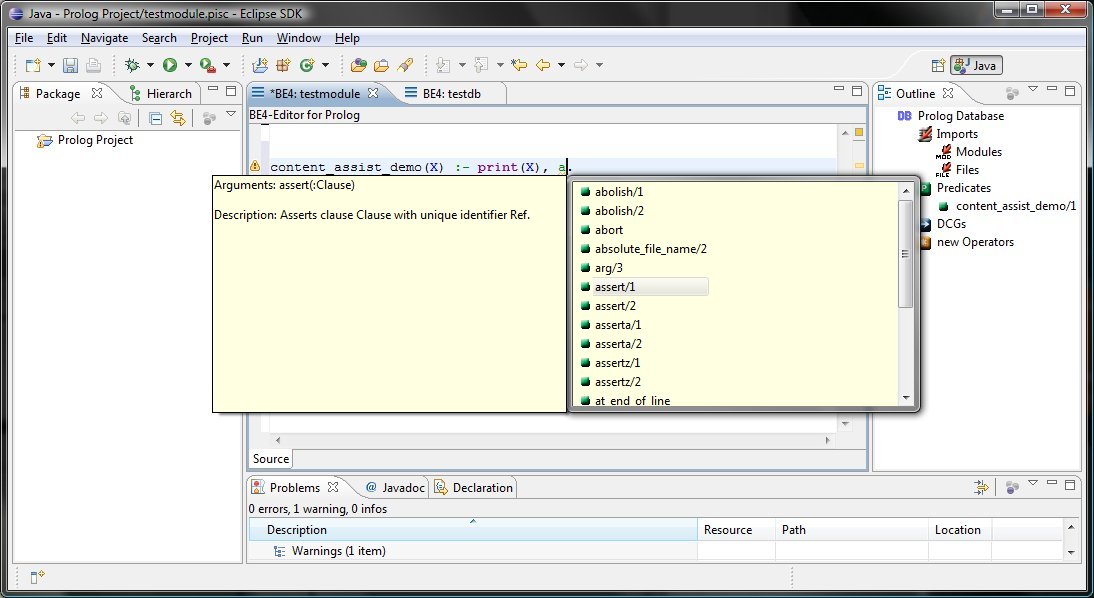}
\centering
\caption{ProClipse Screenshot - Content Assist}
\label{content-assist}
\end{figure}

\paragraph{Content Assist:} To improve faster coding a content assist has been 
implemented which offers content sensitive proposals to automatically complete the 
word the developer types. As can be seen in Figure~\ref{content-assist},
  content assist can also be used to retrieve 
information about predicates, like their synopsis and usage, and present an overview
of the available predicates, modules, DCGs and more.

\begin{figure}[h]
\centering
\includegraphics[scale=0.31]{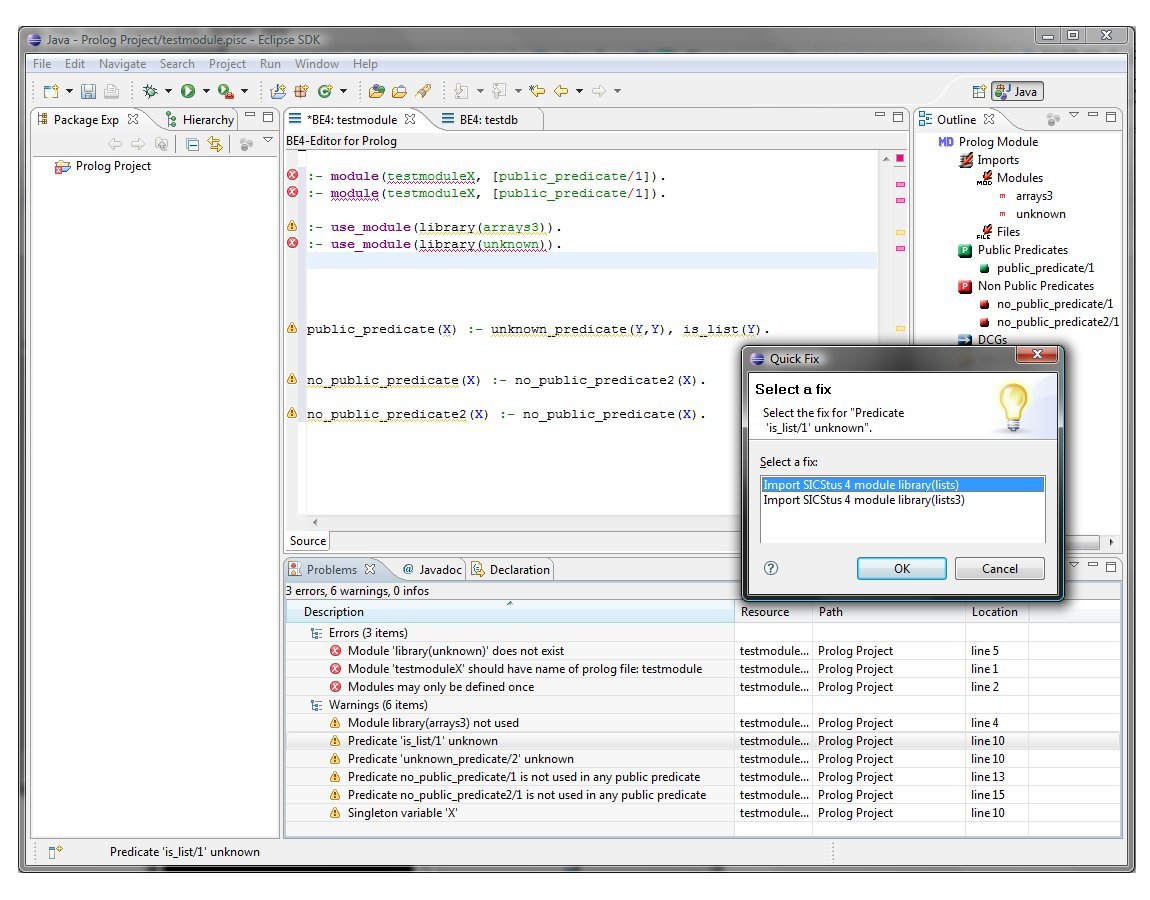}
\centering
\caption{ProClipse Screenshot - Error presentation and Quick Fix}
\label{quickfix}
\end{figure}

\paragraph{Quick Fixes:} Quick fixes offer the possibility to directly auto-correct
erroneous source code via a mouse click. Based on the type of error, a suitable set of
fixes is offered\footnote{Quick fixes can be invoked by right clicking the error in 
the problems view and selecting the "Quick Fix" entry}. For instance, if the 
developer calls an unknown predicate (which is presented as an error in the editor 
view), ProClipse will look for Prolog modules and databases exporting this 
predicate and offer to import one of these. This is illustrated in Figure~\ref{quickfix}.

\begin{figure}[h]
\centering
\includegraphics[scale=0.31]{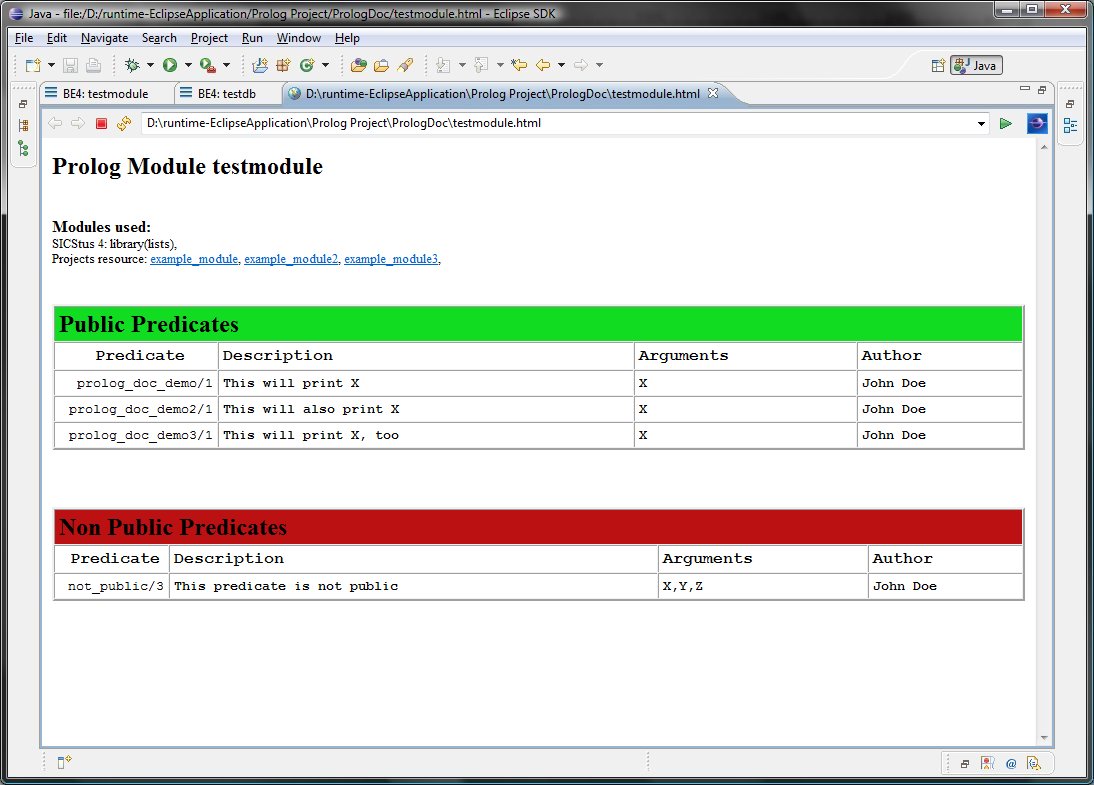}
\centering
\caption{ProClipse Screenshot - PrologDoc Summary}
\label{PrologDoc}
\end{figure}

\paragraph{PrologDoc:} Inspired by the documentation generator JavaDoc from Sun Microsystems for Java, we created PrologDoc. To attach a documentation to a predicate, a comment field containing PrologDoc entries\footnote{Default PrologDoc entries are: 'Author:', 'Arguments:', 'Description:'.} must be written above the first appearance of a clause defining this predicate. Documentation can also be attached to entire modules, 
 in which case the PrologDoc comment field must be written above the module definition. 
Our implementation can also extract a PrologDoc summary for a complete project.
This generated summary is presented in HTML format (see Figure~\ref{PrologDoc}), with the ability to directly 
 navigate into the summaries of imported Prolog modules or files. It shows an outline of all defined predicates, including their synopsis.

\paragraph{Text Hover:} ProClipse offers two types of text hover which can deliver quick information of a 
 lexical token in our editor view. For instance, if a developer wants to know how a 
certain predicate has been defined, he simply has to point the mouse at the predicate. The text hover can also present the synopsis and the arguments of a Prolog built-in, the exported predicates of  imported Prolog modules or files, and the definition of an user-defined operator.
By additionally pressing the shift key the PrologDoc hover is shown, which displays the PrologDoc entries of a predicate or module.


\section{Architecture and Implementation}

\subsection{BE4 Phases}
Our development is based on BE4 \cite{BeLeu_B07}, which contains
 a multi-phase parsing and analysis framework.
The architecture was designed for extensibility by new, as of yet unknown, plug-ins.

\begin{figure}[h]
	\begin{center}
	\includegraphics[width=12cm]{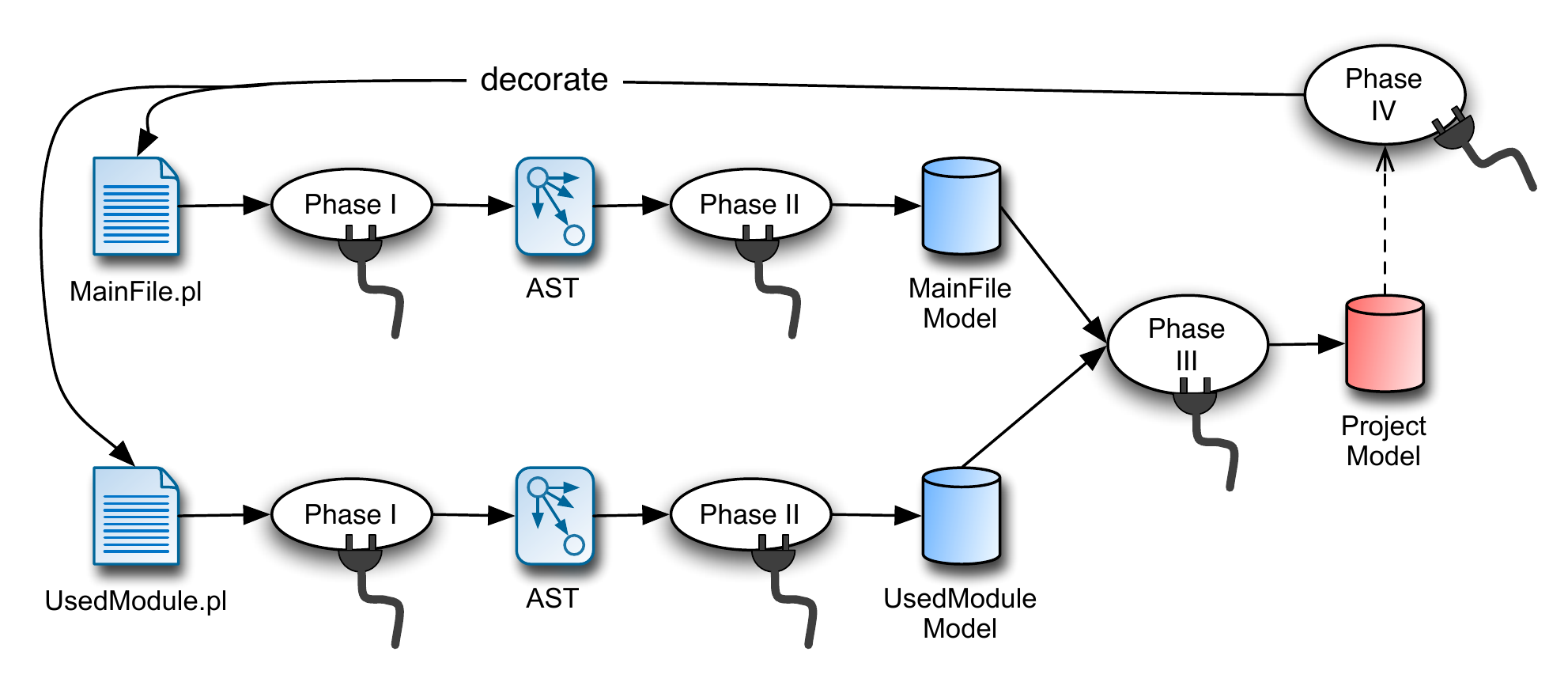}
	\end{center}
   \caption{Phases of the building framework} 
\label{builder_details}
\end{figure}

As shown in Figure~\ref{builder_details}, our tool is decomposed into four phases.

\begin{itemize}
\item
The first phase is the parsing phase which constructs an abstract syntax tree (AST). To that end we have
  written a Parser and Engine for full Prolog in Java (see Subsection~\ref{sub:parser} below).
 
\item
Phases II and III decorate this AST and then combine various ASTs from multiple files to perform global analysis.

\item
The last phase (IV) generates the relevant information for the semantics aware editor to work and decorates the source files,
 e.g., generating Eclipse markers\footnote{{\tt\scriptsize http://www.eclipse.org/articles/Article-Mark My Words/mark-my-words.html}}.
\end{itemize}

\subsection{Parser}
\label{sub:parser}
The centerpiece of a semantic-aware editor for programming languages is a parser that generates a model from source-code. 
In order to provide seamless integration into Eclipse and BE4, we have written a Prolog parser in Java.

Our parser generates a fully typed%
\footnote{Note that we do type the individual nodes in the abstract syntax tree, but we do not try to infer
 Prolog types for arguments of predicates.}
 parse tree of the processed Prolog code.
Note that in order to parse full Prolog with directives and operator declarations, a Prolog engine is required.
We have hence also developed a simple Prolog engine in Java, 
 with  support for a basic set of built-ins.
Our parser and engine fully support the dynamic operator definitions and the dynamic grammar associated to them.
Our Prolog parsing framework provides a mechanism to attach custom post-processing steps to the results produced by the parser, so they can be used in a flexible way in different contexts.

The Prolog parser is built of a set of different components which also represent the different steps of the evaluation process. These are the lexical analysis, the parser and a series of post-processing steps called engines.

\subsubsection{Lexical analysis} 
\label{ssub:lexical_analysis}
The lexer implemented for Prolog was created using JFLex and is based on the definition of the language as provided in the SICStus Prolog User's Manual \cite{SICStus2007}. Every token type is represented by an instance of a corresponding Java class and holds information about the
 source text it matched and where on the input stream it appeared.

Prolog tokens can have different meanings depending on their position in the input stream and the state of the program 
 (itself depending on operations performed in previous steps). The lexer takes this information into account when generating lexical tokens, performing a state aware token generation, so that the same text can be represented by different tokens depending on the context.
 E.g. the '+' atom would be represented as an atom or an operator, depending of the current operator definitions stored in the internal database.

\subsubsection{Syntactic Analysis} 
\label{sub:syntactic_analysis}
The parser processes the token stream on a sentence by sentence basis. Every time the abstract syntax tree for a sentence is constructed, it is dispatched to the post-processing steps for evaluation, before resuming the parsing process. This way the parser can take into account possible modifications to the environment done by the evaluation of the preceding sentence, such as an operator definition.

The algorithm used  in the parser follows the shift reduce parsing principle, with one token lookahead \cite{Aho:Dragon2}. Instead of using an approach based on DFA as described in \cite{Appel2002}, this parser takes advantage of the simple structure of the Prolog grammar, which makes it unnecessary to track multiple possible productions for a given input at the same time.
Our algorithm is based on the current token and as necessary on the previous and next tokens on the input stream. 

This approach allows to provide a fine-grained and fully typed AST, providing as much information as possible about the language tokens in the context they appeared.

\subsubsection{Post-Processing, The Prolog Runtime} 
\label{ssub:Prolog_runtime}
After a sentence has been parsed, the AST representing it is dispatched to the post-processor. The post-processor is built of a chain of objects called engines. These engines are responsible of performing different analysis and execution steps on the AST and to pass the AST to next engine registered in the chain. There are different engines with different responsibilities, such as executing the code, storing it in the internal term database or providing an interactive read eval loop as a command line interface.
There is a special engine responsible for evaluation directives, every time a directive passes this engine the directive is executed, so additional files get loaded and modifications to internal settings are executed.


\subsubsection{Error Recovery}
To provide as much information as possible in case of an error, the parser supports a simple form error recovery by dropping tokens until it can resume the parsing process normally. All errors are collected and made available to the user, so that they can be used - for instance - to highlight the errors in an IDE. 
This enables our plugin to detect multiple errors in a source file (and not just the first error).

\section{Related Work and Conclusion} \label{related}

Various Prolog systems come with some support for convenient editing and development.
For example, SICStus Prolog and Ciao Prolog are distributed with an Emacs-mode for Prolog.
SWI Prolog has an Emacs clone based on its XPCE package as well as an editor for Windows.
Visual Prolog and LPA Prolog both have a custom editor for Windows.

Other related work is \cite{SerebrenikEtAl:TPLP08} which present a refactoring tool for Prolog based on the editor VIM.
In future we plan to add refactoring capabilities to our plugin.

While existing Prolog documentation tools such as PrologDoc\footnote{http://prologdoc.sourceforge.net/} provide a wide range of documentation options, our tool currently only provies a subset of these options, but has the advantage of beeing fully integrated whith eclipse. In future we plan to integrate PrologDoc with our documentation tools to allow more detailed documentation.

On the Eclipse side we are aware of the following three plugins.

\begin{itemize}
\item {\bf Prolog Plugin}%
\footnote{http://eclipse.ime.usp.br/projetos/grad/plugin-prolog/index.html}
 developed in 2003/2004.
It is unclear whether this Plugin is still maintained
  and it only seems to provide syntax highlighting and consulting a Prolog interpreter.

\item
 {\bf PDT} %
\footnote{http://roots.iai.uni-bonn.de/research/pdt/}
This plugin supports more features  and links up with SWI Prolog.
Quick fixes, document generation and hover information do not seem to be supported yet.

\item
{\bf ProDevTools}%
\footnote{http://prodevtools.sourceforge.net/} only supports SWI Prolog
  and does not yet seem to support quick fixes and document generation.

\end{itemize}

All three plugins definitely have potential.
One distinguishing aspect
 of our plugin is the language-independent BE4 framework, which will be maintained in the foreseeable
  future.
 Improvements made to BE4 for other languages, such as B or CSP, will feed back into the Prolog plugin as well.

We also plan to keep our editor largely independent from any particular Prolog system and plan to integrate
 more analysis information based on abstract interpretation. While it is easy to create analysis tools written in java, in future we will investigate ways of integrating analysis tools written in prolog without having to dissmis the independence from any Prolog system.
 
We also plan to integrate termination analysis and partial evaluation tools into our plugin.

\bibliographystyle{abbrv}

\end{document}